\documentclass[final,5p,times,twocolumn]{elsarticle}
\biboptions{sort&compress}
\usepackage[T1]{fontenc}
\usepackage{amsmath,amssymb,bm,booktabs,graphicx}
\usepackage{microtype}
\usepackage{multirow}
\usepackage{xcolor}

\usepackage[colorlinks=true,linkcolor=blue,citecolor=blue,urlcolor=blue]{hyperref}
\usepackage{placeins}

\journal{Physics Letters B}

\newcommand{\dd}{\mathrm{d}}
\newcommand{\Id}{\mathbf{I}}
\hypersetup{pdftitle={Reassessing $X(2370)$ in $J/\psi\to\gamma K_S^0K_S^0\pi^0$ beyond the one-dimensional mass projections},pdfauthor={Xiang-Kun Dong, Teng Ji, Meike K\"u{\ss}ner, Ulf-G. Mei\ss ner},pdfsubject={working manuscript}}
\begin{document}
\begin{frontmatter}
\title{Reassessing the $X(2370)$ in $J/\psi\to\gamma K_S^0K_S^0\pi^0$ beyond the one-dimensional mass projections}

\author[a,b]{Xiang-Kun Dong}
\ead{xiangkun@hiskp.uni-bonn.de}
\author[a,b]{Teng Ji}
\ead{teng@hiskp.uni-bonn.de}
\author[c]{Meike K\"u{\ss}ner}
\ead{ mkusner@wm.edu}
\author[a,b,d,e]{Ulf-G. Mei\ss ner}
\ead{meissner@hiskp.uni-bonn.de}

\affiliation[a]{organization={Helmholtz-Institut f\"ur Strahlen- und Kernphysik and Cluster of Excellence ``Color meets Flavor'', Universit\"at Bonn},
                city={Bonn}, postcode={D-53115}, country={Germany}}
\affiliation[b]{organization={Bethe Center for Theoretical Physics, Universit\"at Bonn},
                city={Bonn}, postcode={D-53115}, country={Germany}}
\affiliation[c]{organization={School of Computing, Data Sciences \& Physics, College of William \& Mary},
                city={Williamsburg}, postcode={VA 23187-8795}, country={USA}}
\affiliation[d]{organization={Institute for Advanced Simulation (IAS-4) and Cluster of Excellence ``Color meets Flavor'', Forschungszentrum J\"ulich},
                city={J\"ulich}, postcode={D-52425}, country={Germany}}
\affiliation[e]{organization={Peng Huanwu Collaborative Center for Research and Education, International Institute for Interdisciplinary and Frontiers, Beihang University},
                city={Beijing}, postcode={100191}, country={China}}

\begin{abstract}
	The BESIII Collaboration recently extracted the properties of the $X(2370)$ from a fit to the one-dimensional $K_S^0K_S^0\pi^0$ invariant-mass spectrum in $J/\psi\to\gamma K_S^0K_S^0\pi^0$. The intermediate two-body structures visible in the mass correlations, including the $K_0^*(1430)$ bands, require a coherent treatment of cross-channel interference when interpreting the enhancement near 2.3~GeV. We reassess the need for an additional $X(2370)$ contribution by fitting the published data using an effective coupled-channel framework with an approximate detection efficiency depending on the three-body invariant mass.  The three-body parent amplitude is parameterized using a $K$-matrix that describes scattering among several quasi-two-body channels. The seven- and eight-bare-state models give similar mass distributions. Adding the eighth bare state lowers the Poisson deviance by 1.22\% and produces an additional pole at $E=(2419-i\,152)$~MeV, substantially broader than the $X(2370)$ mass--width reference. This pole is model-dependent, and its identification with the $X(2370)$ is not clear. The present analysis does not establish the need for an additional $X(2370)$ contribution in $J/\psi\to\gamma K_S^0K_S^0\pi^0$, thereby weakening the flavor-singlet argument based on the suppression of $K^*\bar K$ relative to the total $K\bar K\pi$ contribution. A more comprehensive experimental amplitude analysis, incorporating coherent interference, mass and angular distributions, and detector effects across related decay channels, is needed to establish the role and decay properties of the $X(2370)$.
\end{abstract}
\end{frontmatter}
\raggedbottom

\section{Introduction}\label{sec:intro}
The self-interaction of gluons in quantum chromodynamics (QCD) allows the formation of color-singlet gluonic bound states, or glueballs, whose spectroscopy has been studied since the pioneering works~\cite{Fritzsch:1975tx,Jaffe:1975fd}. Lattice calculations provide quantitative predictions for the pure-gauge spectrum~\cite{Morningstar:1999rf,Chen:2005mg}, but relating these states to physical hadrons requires accounting for dynamical quarks. In full QCD, quark--antiquark pair creation allows gluonic configurations to couple to states containing quarks and to hadronic decay channels. Consequently, substantial quark and multiquark components cannot be excluded for a resonance in this mass region, and proximity to a pure-gauge mass prediction does not determine its composition. We refer to Refs.~\cite{Mathieu:2008me,Vadacchino:2023vnc,Chevalier:2026pwf} for related reviews. 

Over the past fifteen years, BESIII has progressively established the experimental properties of the $X(2370)$. It was first observed in the $\pi^+\pi^-\eta'$ invariant-mass spectrum of $J/\psi\to\gamma\pi^+\pi^-\eta'$ in 2011~\cite{BESIII:2010gmv}, and subsequently confirmed in both charged and neutral $K\bar K\eta'$ modes~\cite{BESIII:2019wkp}. Its observation in the electromagnetic Dalitz decay $J/\psi\to e^+e^-X(2370)$ extended the evidence to production through a virtual photon~\cite{BESIII:2021xoh}. A partial-wave analysis of $J/\psi\to\gamma K_S^0K_S^0\eta'$ determined its quantum numbers to be $J^{PC}=0^{-+}$~\cite{BESIII:2023wfi}.

The growing set of decay measurements provides further constraints. No significant $X(2370)$ signal was found in $\eta\eta\eta'$~\cite{BESIII:2020ymv}, and a partial-wave analysis of $J/\psi\to\gamma\gamma\phi$ placed an upper limit on its decay into $\gamma\phi$~\cite{BESIII:2024ein}. More recently, BESIII reported the $K_S^0K_S^0\pi^0$, $\pi^0\pi^0\eta$, and $a_0(980)\pi$ modes, emphasizing their similarity to the decay pattern of $\eta_c$~\cite{BESIII:2026rzt}. A subsequent search found no significant $X(2370)\to K^*(892)\bar K$ contribution and interpreted this suppression as evidence for a flavor-singlet assignment~\cite{BESIII:2026mvn}. BESIII interpreted these observations, together with the mass, quantum numbers, and production and decay properties, as supporting a dominant pseudoscalar-glueball component~\cite{BESIII:2026mvn}. This interpretation requires distinguishing gluonic dominance from other configurations that can share the same quantum numbers and flavor properties.

The theoretical literature offers several competing descriptions. Conventional pseudoscalar excitations and multiquark configurations have long been considered~\cite{Yu:2011ta,Deng:2012wi,Wang:2025nme}. An analysis of the earlier branching-fraction data found tension between the inferred radiative production rate and the lattice prediction for a pure glueball~\cite{Sun:2021kka}. More recent studies obtain an enhanced rate by introducing glueball--charmonium mixing~\cite{Chen:2026lki,Zhu:2026zju}. Glueball-based descriptions of the decay patterns~\cite{Tan:2026uue,Tan:2026crv,Wang:2026iyq,Giacosa:2026pqh}, mass and width~\cite{Buisseret:2026acu}, and radial assignment~\cite{Li:2024fko} have also been proposed, with conclusions that depend on the assumed dynamics. The recent flavor argument has also prompted alternatives: a flavor-singlet radial hybrid can accommodate the suppressed $K^*\bar K$ mode~\cite{Lu:2026foy}, and comparisons of three-pseudoscalar decay patterns leave several flavor-singlet configurations viable~\cite{Zhang:2026poz}. A separate study generates a state through $\phi h_1(1380)$ dynamics and relates the enhanced non-$K^*\bar K$ contribution to $K\bar K\pi$ to triangle dynamics~\cite{Khemchandani:2026syv}. The existing phenomenology therefore does not uniquely determine the gluonic content of the $X(2370)$. Before drawing conclusions about its composition, the decay properties used to discriminate among these descriptions must themselves be reliably extracted.

Baryon--antibaryon dynamics provides further motivation for alternatives to a predominantly gluonic state. Pronounced enhancements and threshold structures near the $N\bar N$ threshold have been reported in the $p\bar p$ invariant-mass spectrum~\cite{BES:2003aic,CLEO:2010fre,BESIII:2010vwa,BESIII:2011aa} and in several light-meson production channels~\cite{BES:2005ega,BESIII:2023vvr,BESIII:2016fbr,BESIII:2013sbm,BESIII:2010gmv}. These observations have motivated interpretations of the $X(1835/1880)$ in terms of $N\bar N$ interactions~\cite{Zou:2003zn,Kang:2015yka,Yang:2022kpm,Haidenbauer:2015yka,Niu:2024cfn,Dai:2018tlc,Ji:2026mfe}. Further motivation comes from the $\eta'\pi^+\pi^-$ spectrum in radiative $J/\psi$ decays~\cite{BESIII:2022sfx}, where most of the peaks and dips between 1.8 and 2.8~GeV lie near the thresholds of the four baryon--antibaryon channels. Clear enhancements have also been observed near the $\Lambda\bar\Lambda$ threshold~\cite{BESIII:2021fqx,BESIII:2022tvj,BESIII:2026btb}. Taken together, these observations make it natural to expect analogous near-threshold states in other baryon--antibaryon channels. A meson-exchange model admits an interpretation of $\eta(2225)$~\cite{BESIII:2016qzq} as a spin-singlet $S$-wave $\Lambda\bar\Lambda$ bound state~\cite{Zhao:2013ffn}. More recently, Ref.~\cite{Ben:2026afy} proposed a predominantly $\Sigma\bar\Sigma$ molecular interpretation of $X(2370)$, motivated by its proximity to the threshold near 2.38~GeV and its pseudoscalar quantum numbers, and suggested $X(2600)$ as a possible $\Xi\bar\Xi$ counterpart. These studies motivate explicit baryon--antibaryon coupled-channel tests while leaving the existence and molecular content of the proposed poles to be checked.

The $K\bar K\pi$ signal extractions underlying the flavor-singlet argument rely on one-dimensional mass-spectrum fits~\cite{BESIII:2026rzt,BESIII:2026mvn}. These fits do not describe the coherent interference among the intermediate channels visible in the published two-dimensional distributions, notably the overlapping $K_0^*(1430)$ bands. Such interference can reshape the projected spectrum and alter the inferred resonance yields. In addition, the mass of the $X(2370)$ extracted from $K_S^0K_S^0\pi^0$ differs appreciably from that extracted from other channels~\cite{BESIII:2026rzt}.  The mass correlations must therefore be analyzed before these yields can reliably constrain the properties of the $X(2370)$. Ultimately, a complete amplitude analysis requires the full phase-space information, including angular distributions, together with the detector response. Earlier partial-wave analysis of $J/\psi\to\gamma K^+K^-\pi^0$ by BES~\cite{BES:1998bgh} and $J/\psi\to\gamma K_S^0K_S^0\pi^0$ by BESIII~\cite{BESIII:2022chl} in low energy region provide examples of such approach, followed by a more sophisticated three-body unitary coupled-channel description of the $\eta(1405/1475)$ region~\cite{Nakamura:2022rdd,Nakamura:2023hbt}.

To address this limitation with the available information, we fit digitized versions of the published two-dimensional distributions and the full parent-mass spectrum using an effective dispersive coupled-channel amplitude that treats the intermediate structures and their interference coherently. We compare a seven-bare-state model, $H_0$, with an extension, $H_1$, containing an additional bare state initialized near 2.37~GeV, to assess whether this contribution is needed. The public data nevertheless limit the scope of this test. The two-dimensional intensities are reconstructed from a finite color palette, so their bin contents are approximate. The full phase-space information, including angular distributions, and multidimensional detector response are unavailable. Moreover, the distributions share events, and the statistical correlations cannot be recovered from the published plots. The reconstructed distributions therefore constrain the mass-correlation patterns and allow a descriptive comparison of the models, but do not provide a calibrated experimental likelihood. We consequently do not assign a statistical significance to the additional contribution or quote uncertainties on the fitted resonance parameters and pole positions. These quantities require an experimental amplitude analysis with the original event sample and detector response.

\section{Coupled-channel amplitudes and coherent production}\label{sec:formalism}
\subsection{Effective channels and scattering amplitudes}
Several candidate states populate the parent-mass region, many with large widths and overlapping decay modes. A coherent treatment must therefore account for both their production and the rescattering through common channels. We use a $K$-matrix representation in the isoscalar pseudoscalar sector, restricting the three-body system to $J^{PC}=0^{-+}$. Contributions with other quantum numbers, such as $1^{++}$ and $2^{-+}$, are not included. The published mass distributions do not provide the angular information needed for a reliable separation of these contributions, so this restriction is a model assumption rather than an exclusion of other partial waves. The construction proceeds in three steps: define the coupled quasi-two-body channels, dress their production amplitudes by rescattering, and attach their decays to the observed $K\bar K\pi$ final state.

The effective channels are motivated by the pair structures in the measured distributions. For $a_0(980)\pi$, we use two scalar $K\bar K$ source functions proportional to $T^a_{21}$ and $T^a_{22}$ of the coupled $\pi\eta$-$K\bar K$ amplitude adapted from Ref.~\cite{Albaladejo:2015aca}, with the $\delta_{12}=100^\circ$ parameter set. Here, channels~1 and 2 denote $\pi\eta$ and $K\bar K$, respectively. These functions describe two ways of producing the same interacting scalar subsystem with a spectator pion.

The other resonant configurations are $K^*(892)\bar K$, $K_0^*(1430)\bar K$, $K_0^*(1950)\bar K$, $K_2^*(1430)\bar K$, $K^*(1410)\bar K$, $a_0(1710)\pi$, and $a_2(1320)\pi$. Their pair amplitudes are represented by Breit--Wigner forms with mass-dependent widths. We also include smooth $K\pi$ $S$-, $P$-, and $D$-wave configurations with a spectator kaon. Together, the two scalar source functions, seven Breit--Wigner terms, and three smooth terms give $n=12$ components. The smooth waves are treated on the same footing as the resonant configurations in the coupled-channel construction. \ref{app:shapes} collects the pair amplitudes, angular vertices, and barrier factors.

Only mesonic channels are included in the present model. We do not investigate baryon--antibaryon thresholds, so this analysis does not test the baryon--antibaryon interpretations discussed in the Introduction. This is left for future work.\par

Here, a channel labels an amplitude component feeding the common three-body final state, rather than a distinct stable two-body asymptotic state. Writing $s=M^2$ for the squared three-body invariant mass, we construct the $n\times n$ interaction and scattering matrices as~\cite{Kopf:2020yoa,CrystalBarrel:2019zqh}
\begin{align}
K_{ij}(s)&=B_{ij}+\sum_{\alpha=1}^{N_b}
\frac{g_{i\alpha}g_{j\alpha}}{d_\alpha(s)}~,\nonumber\\
T(s)&=[\Id-K(s)C(s)]^{-1}K(s)~.
\label{eq:scattering}
\end{align}
The real symmetric constant matrix $B$ describes a smooth interaction, and the real couplings $g_{i\alpha}$ connect channel $i$ to bare parent state $\alpha$. The $N_b$ bare states have denominators
\begin{equation}
d_\alpha(s)=M_{0\alpha}^2-s-iM_{0\alpha}^{}\Gamma_{0\alpha}^{}.
\label{eq:bare}
\end{equation}
The nonnegative constant $\Gamma_{0\alpha}$ models absorption into omitted channels. The dressing by the explicitly included channels is generated separately by the loop matrix $C$.

To construct $C$, we associate each channel with a three-body decay function $f_i$ containing the pair line shapes and the angular factors appropriate to a pseudoscalar parent. Their explicit forms are given in \ref{app:shapes} and summarized in Table~\ref{tab:basis}. The spectral matrix $\rho_{ij}$ is obtained by integrating the overlap $f_i^*f_j$ over three-body phase space. The spectral density and its multiplicity convention are specified in \ref{app:spectral}. The corresponding once-subtracted dispersion relation is
\begin{equation}
C(z)=\frac{z}{\pi}\int_{s_{\rm th}}^\infty
\frac{\rho(s')}{s'(s'-z)}\dd s',\qquad C(0)=0~,
\label{eq:dispersion}
\end{equation}
where $s_{\rm th}=(2M_K+M_\pi)^2$. This follows the dispersive construction of Refs.~\cite{Paris:2010tz,Workman:2012jf}. On the upper rim of the physical cut, the absorptive part satisfies $(C_+-C_+^\dagger)/(2i)=\rho$. Thus the same decay functions determine both the rescattering matrix and the final-state amplitude below. This construction implements coupled-channel unitarity constraints within the chosen channel space, supplemented by phenomenological omitted-channel absorption. However, it does not constitute a complete treatment of three-body unitarity.

The baseline model in this work, $H_0$, contains $N_b=7$ bare parent states initialized near the $\eta(1295)$, $\eta(1405)$, $\eta(1475)$, $X(1835)$, $X(2120)$, $X(2600)$, and $\eta_c(1S)$. All terms are assigned to the $I=0$, $J^{PC}=0^{-+}$ sector considered here. These assignments define the model and do not imply experimentally established quantum numbers for every candidate. The bare masses and couplings are fitted, so the state labels specify the initialization rather than identify the poles of the dressed amplitude. To test the need for an additional contribution near 2.37~GeV, we also consider $H_1$, which adds one bare state initialized at this mass, giving $N_b=8$. The two hypotheses use the same twelve-channel space and the same production and rescattering framework. Their fitted parameters and resulting amplitudes are compared in Sec.~\ref{sec:results}.

\subsection{Source production and rescattering}
The radiative source can produce each configuration directly or through a bare parent state. Before rescattering, these contributions form the $n$-component production vector
\begin{equation}
P_i(s)=p_i(s)+\sum_{\alpha=1}^{N_b}
\frac{g_{i\alpha}\beta_\alpha}{d_\alpha(s)}~,
\label{eq:source}
\end{equation}
where $p_i(s)$ is a smooth function taken to be linear in $s$, and $\beta_\alpha$ is the complex bare-state production coupling. The linear function also has complex coefficients. The couplings $g_{i\alpha}$ and denominators $d_\alpha$ are the same as in the scattering matrix, so production and rescattering share the same bare-state structure.

The full production amplitude follows the standard $P$-vector prescription of the $K$-matrix formalism~\cite{Aitchison:1972ay,Chung:1995dx},
\begin{equation}
F(s)=[\Id-K(s)C(s)]^{-1}P(s)~,
\label{eq:kp}
\end{equation}
with the dispersive loop matrix $C$ defined in Eq.~\eqref{eq:dispersion}.

\subsection{Coherent three-body decay amplitude}
The vector $F$ describes production of the coupled configurations. Attaching their decay functions gives the amplitude for the observed final state,
\begin{equation}
\mathcal A(s,t,u)=\sum_{i=1}^{n}F_i(s)f_i(s,t,u),
\label{eq:threebody}
\end{equation}
where $t=(p_1+p_\pi)^2$ and $u=(p_2+p_\pi)^2$, with $p_1,p_2$ the two kaon momenta. 
For $K_S^0K_S^0\pi^0$, each $K\pi$ component contains the coherent sum of the two kaon assignments. Squaring Eq.~\eqref{eq:threebody} therefore keeps interference both within a given component and between different components, as well as between the parent contributions and smooth production contained in $F$. In particular, the overlapping $K_0^*(1430)$ bands contribute through their amplitude sum, not through an incoherent sum of intensities. This interference links the structures in the two-dimensional distributions to their one-dimensional projections.

The coherent decay intensity is
\begin{equation}
\dd\Gamma\propto E_\gamma^3\,|\mathcal A(s,t,u)|^2\,\dd s\,\dd\Phi_3,
\label{eq:intensity}
\end{equation}
where $E_\gamma$ is the photon energy in the $J/\psi$ rest frame and $\dd\Phi_3$ is the three-body phase-space element. For the observed distributions, the intensity in Eq.~\eqref{eq:intensity} is multiplied by the efficiency $\varepsilon(M)$ before integration over each bin. We approximate $\varepsilon(M)$ by a fixed function of the three-body invariant mass, inferred from the published reconstructed phase-space simulation, while neglecting its two-body-mass and angular dependence and any explicit resolution convolution. The construction and its limitations are given in \ref{app:efficiency}.  The same amplitude parameters describe both two-dimensional distributions and the full parent-mass spectrum, without an independent normalization for the latter.

The same observed final state can also arise from $J/\psi\to K^{*0}\bar K^{*0}$, with one vector kaon decaying radiatively into $\gamma K^0$ or $\gamma\bar K^0$ and the other into $\bar K^0\pi^0$ or $K^0\pi^0$, followed by reconstruction of both neutral kaons as $K_S^0$. This production mechanism is not included in our amplitude and is distinct from the $K^*\bar K$ subchannels of the radiatively produced pseudoscalar system. Its contribution would require a dedicated assessment in an experimental amplitude analysis.\par

\section{Data representation and comparison protocol}\label{sec:data}
We reconstruct the two-dimensional mass distributions in Fig.~2 of the BESIII preprint~\cite{BESIII:2026rzt} by mapping the displayed bin colors to the published color scale. The resulting $D_{ab}^{(r)}$ estimate the displayed intensities. We also digitize the complete published one-dimensional parent-mass spectrum over approximately 1.15--3.10~GeV, which supplies the low-mass information absent from the two-dimensional images, whose displayed data start near 1.6~GeV. With the limitations discussed in the Introduction, these inputs are used for a descriptive model comparison.

Integrating $\varepsilon(M)$ times Eq.~\eqref{eq:intensity} over the physical region of each experimental bin gives the predictions used in the fit, denoted by $\mu_{ij}^{(r)}$, with $r=KK,K\pi$ and $KK\pi$. Let $D_{ij}^{(r)}$ denote the color-scale-reconstructed intensity. We define the Poisson deviance~\cite{BakerCousins1984} per bin by
\begin{align}
d_{ij}^{(r)}&=2\left[\mu_{ij}^{(r)}-D_{ij}^{(r)}+D_{ij}^{(r)}\ln\frac{D_{ij}^{(r)}}{\mu_{ij}^{(r)}}\right].
\label{eq:deviance}
\end{align}
The fitted objective is
\begin{equation}
D=\frac12\sum_{(i,j)\in\Omega_{KK}}d_{ij}^{(KK)}
+\frac12\sum_{(i,j)\in\Omega_{K\pi}}d_{ij}^{(K\pi)}
+\sum_{a=1}^{65}d_a^{(KK\pi)},
\label{eq:objective}
\end{equation}
where $\Omega_r$ contains the bins intersecting the kinematically allowed region. The fit includes 6264 $KK$ bins and 6006 $K\pi$ bins. The parent contribution uses the same deviance with the digitized parent spectrum and its bin-integrated prediction. Each two-dimensional distribution carries weight $1/2$, and the parent spectrum carries weight one. These weights define a descriptive objective, not a treatment of the correlations among the three distributions.

\section{Results and discussion}\label{sec:results}

\begin{figure}[!tbp]
\includegraphics[width=\linewidth]{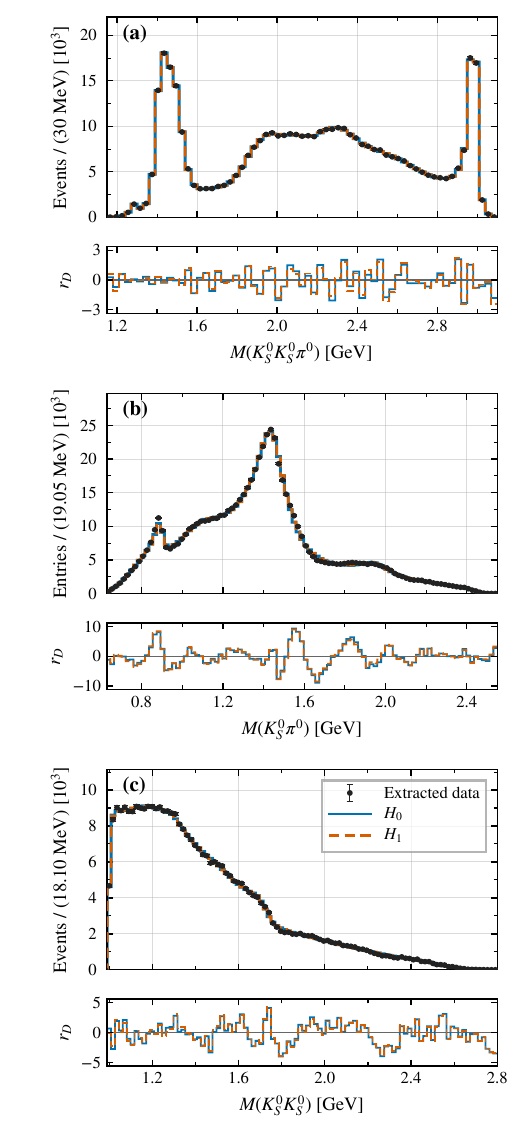}
\caption{Mass distributions for $J/\psi\to\gamma K_S^0K_S^0\pi^0$: (a) the three-body invariant mass, (b) the $K_S^0\pi^0$ invariant mass, and (c) the $K_S^0K_S^0$ invariant mass. Black points denote digitized data with error bars that are barely visible. The blue solid curves show $H_0$, and the orange dashed curves show $H_1$. The pair-mass distributions are projections of the reconstructed color-scale intensities, rather than the original event counts. Both $K_S^0\pi^0$ combinations are counted in panel (b). The lower panels show the signed Poisson-deviance residuals $r_D=\operatorname{sgn}(N-\mu)\sqrt{2[\mu-N+N\ln(N/\mu)]}$ for the corresponding projected data and predictions.}\label{fig:projections}
\end{figure}

Table~\ref{tab:fit} summarizes the fits obtained by minimizing the descriptive Poisson-deviance objective in Eq.~\eqref{eq:objective}. Adding the eighth bare term reduces $D$ by 96.4 (1.22\%), while the projected line shapes are almost indistinguishable in the two models, as shown in Fig.~\ref{fig:projections}. The corresponding two-dimensional comparisons and residual maps are shown in \ref{app:numerical}. The main difference between the two descriptions concerns the pole content associated with these distributions. A coherent decomposition of the $H_0$ parent-mass spectrum and the origin of its enhancement near $2.37\,\mathrm{GeV}$ are discussed in Appendix~\ref{app:decomposition}.

The improvement of the global objective should not be interpreted as a statistical significance for an additional resonance. The fitted inputs are digitized mass projections with shared events and unavailable correlations, which preclude a calibrated likelihood-ratio interpretation. Moreover, much of the improvement arises outside the $2.37\,\mathrm{GeV}$ region, whose structure is already reproduced by $H_0$. Our comparison therefore tests the ambiguity of the resonance interpretation, rather than establishing a statistical preference for either model or excluding the $X(2370)$.

We tested the sensitivity to digitization by assigning intensities to the lower and upper boundaries of each color interval, setting the lowest color level to zero, and sampling intensities uniformly within their color intervals. These variations leave the description of the enhancement near $2.37\,\mathrm{GeV}$ qualitatively unchanged, while the broader poles show larger shifts than the narrower ones, reflecting their weaker constraints from the available mass projections.

\begin{figure}[!tbp]
\centering
\includegraphics[width=\columnwidth]{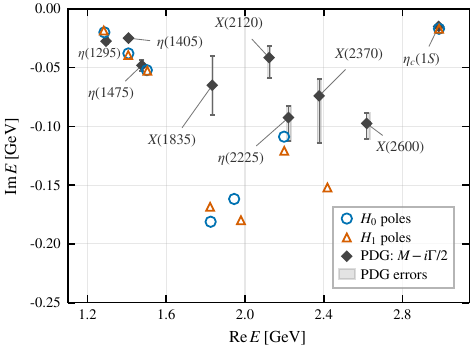}
\caption{Pole positions for the best-fit $H_0$ (blue circles) and $H_1$ (orange triangles) models within $-0.25<\operatorname{Im}E<0$ GeV. Black diamonds denote selected PDG 2026 mass--width references at $M-i\Gamma/2$~\cite{ParticleDataGroup:2026mpi}, with uncertainties marked by gray ranges. }\label{fig:continued_poles}
\end{figure}

The pole content is shown in Fig.~\ref{fig:continued_poles} and Table~\ref{tab:poles} for $-0.25<\operatorname{Im}E<0$~GeV. The continuation through the physical cut is defined by $C_{\rm cont}(z)=C_{\rm I}(z)+2i\rho(z)$, with $\rho(z)$ continued vertically from its real-axis anchor and its branches followed along the path. We identify seven poles for $H_0$ and eight for $H_1$.

The four relatively narrow poles in the $\eta(1295)$, $\eta(1405)$, $\eta(1475)$ and $\eta_c(1S)$ regions change little between the two models. The broader structures are more model-dependent. Both models contain a broad pole near 1.83~GeV and one near 2.20~GeV. Additional broad poles lie at $E=(1945-i\,162)$~MeV in $H_0$ and $(1980-i\,180)$~MeV in $H_1$, whose state assignments are left open. Neither model has an identified pole near the $X(2600)$ reference.

Of particular interest is the pole at $E=(2419-i\,152)$~MeV associated with the extra bare state in $H_1$. Its width of 304~MeV is substantially larger than the PDG mass--width reference for the $X(2370)$, and its identification with that state remains uncertain. Together with the modest change in the fitted distributions, this result leaves a distinct $X(2370)$ contribution unestablished in the present analysis.

\begin{table}[t]\centering
\caption{Fits to the full parent-mass spectrum and both two-dimensional mass-correlation distributions. The last three columns give the weighted contributions to $D$ in Eq.~\eqref{eq:objective}. We use the weighted nominal count $\mathrm{dof}=(6264+6006)/2+65-N_{\rm par}$, giving 5987 and 5971 for $H_0$ and $H_1$.}\label{tab:fit}
\setlength{\tabcolsep}{4pt}
\begin{tabular}{ccccccc}\toprule
Model&$N_{\rm par}$&$D$&$D/\mathrm{dof}$&$D^{(KK)}/2$&$D^{(K\pi)}/2$&$D^{(KK\pi)}$\\\midrule
$H_0$&213&7919.1&1.32&3797.2&4053.9&68.0\\
$H_1$&229&7822.7&1.31&3718.8&4033.5&70.4\\
\bottomrule\end{tabular}\end{table}

\begin{table}[t]\centering
\caption{Pole positions of the scattering amplitudes in MeV. PDG entries in terms of $E=M-i\Gamma/2$ are mass--width references~\cite{ParticleDataGroup:2026mpi}, not necessarily pole determinations. The row labels indicate comparison regions, not established state assignments, particularly for the broad poles. A dash means no association is made in the current models.}\label{tab:poles}
\begin{tabular}{lccc}\toprule
Reference & PDG & $H_0$ & $H_1$\\\midrule
$\eta(1295)$ & $1294-i\,28$ & $1286-i\,20$ & $1282-i\,18$\\
$\eta(1405)$ & $1408-i\,25$ & $1407-i\,38$ & $1407-i\,39$\\
$\eta(1475)$ & $1476-i\,48$ & $1501-i\,52$ & $1504-i\,53$\\
$\eta_c(1S)$ & $2984-i\,15$ & $2986-i\,17$ & $2986-i\,17$\\
$X(1835)$ & $1834-i\,65$ & $1826-i\,181$ & $1823-i\,168$\\
Unassigned & --- & $1945-i\,162$ & $1980-i\,180$\\
$\eta(2225)$ & $2221-i\,93$ & $2199-i\,109$ & $2200-i\,121$\\
$X(2370)$ & $2377-i\,74$ & --- & $2419-i\,152$\\
$X(2120)$ & $2122-i\,42$ & --- & ---\\
$X(2600)$ & $2618-i\,98$ & --- & ---\\
\bottomrule\end{tabular}
\end{table}

This uncertainty in the $X(2370)$ contribution affects the flavor-singlet argument, which relies on the suppression of $K^*\bar K$ relative to the total $X(2370)\to K\bar K\pi$ contribution. In the flavor-symmetry limit, the generalized $G$-parity forbids the decay of a $0^{-+}$ flavor singlet into $K^*\bar K$~\cite{Lipkin1981Flavor,Lipkin1982Flavor,Klempt2007Review}. Using this selection rule, BESIII interprets the upper limit $\mathcal R<0.081$ at 90\% confidence level as evidence for a flavor-singlet assignment~\cite{BESIII:2026mvn}, where $\mathcal R$ is the fraction of $X(2370)\to K_S^0K_S^0\pi^0$ proceeding through $K^{*0}\bar K^0+\mathrm{c.c.}$. The ambiguity identified here concerns the denominator of this ratio. The $X(2370)$ yield in the full $K_S^0K_S^0\pi^0$ sample, and hence the inferred $K\bar K\pi$ branching fraction, could be much smaller than that obtained from the one-dimensional fits. If so, the small $K^*\bar K$ yield need not indicate selective suppression within $K\bar K\pi$, and the upper limit on the numerator would no longer establish a small ratio. The present results thus weaken the evidence for selective $K^*\bar K$ suppression used to support the flavor-singlet assignment and, in turn, the glueball interpretation. A small total $K\bar K\pi$ branching fraction would also weaken the proposed analogy with $\eta_c$, for which this is a major decay mode~\cite{ParticleDataGroup:2026mpi}.\par

Resolving this ambiguity requires a common coherent amplitude analysis that extracts both the total $X(2370)\to K\bar K\pi$ contribution and its $K^*\bar K$ component. Such an extraction must test its sensitivity to the model assumptions, including the restriction to the pseudoscalar partial wave, the approximate one-dimensional efficiency, the fixed two-body line shapes, the truncated channel and bare-state content, and the constant phenomenological absorption into omitted channels used here. A simultaneous experimental analysis of $K\bar K\pi$, $K\bar K\eta$, $K\bar K\eta'$, $\eta\pi\pi$, and $\eta'\pi\pi$, using the full event samples and detector responses, would then allow the relative decay strengths to be assessed consistently before drawing conclusions about the flavor structure and gluonic content of the $X(2370)$.

\section{Conclusions}\label{sec:conclusion}
We have reassessed the $K_S^0K_S^0\pi^0$ enhancement in radiative $J/\psi$ decays using the published two-dimensional mass correlations together with the full parent-mass spectrum. Our effective coupled-channel amplitude describes quasi-two-body rescattering and coherent interference under the assumption $J^{PC}=0^{-+}$, with a common approximate parent-mass efficiency applied to all observed distributions. The seven- and eight-bare-state models give similar mass distributions and contain seven and eight poles, respectively, within $-0.25<\operatorname{Im}E<0$~GeV. Adding the eighth bare state lowers the Poisson deviance by 1.22\% and produces a broad pole at $E=(2419-i\,152)$~MeV, whose identification with the $X(2370)$ remains uncertain. The present comparison therefore does not establish the need for a distinct $X(2370)$ contribution in this channel. Only mesonic channels are included. Baryon--antibaryon thresholds and their possible role in the enhancement have not been modeled or tested, and are left for future work.

The uncertainty in the total $X(2370)\to K\bar K\pi$ contribution also affects the interpretation of the small $K^*\bar K$ yield. If the total $K\bar K\pi$ branching fraction is smaller than inferred from one-dimensional fits, the small $K^*\bar K$ yield alone does not establish selective suppression relative to that total. This weakens the corresponding evidence for a flavor-singlet assignment and a dominant glueball component, without excluding either possibility. Testing this argument requires a common coherent extraction of the total $X(2370)$ contribution and its $K^*\bar K$ component. An experimental amplitude analysis with the full event samples and detector responses, extended across related decay channels and allowing for relevant baryon--antibaryon threshold dynamics, is needed to determine the decay pattern and clarify the nature of the $X(2370)$.

\section*{Acknowledgements}
This work is supported in part by the Deutsche Forschungsgemeinschaft (DFG) under Grant No.~525056915 and under Germany's Excellence Strategy---EXC 3107---Project-ID 533766364, by the CAS President's International Fellowship Initiative (PIFI) under Grant No.~2025PD0022,  by the European Research Council (ERC) under the European Union's Horizon 2020 research and innovation programme (EXOTIC, grant agreement No.~101018170) and by the
W\&M Seed Fund.

\appendix

\section{Two-body shapes and angular factors}\label{app:shapes}
We label the two identical neutral kaons by $K_1$ and $K_2$, with four-momenta $p_1$ and $p_2$, respectively, and denote the pion four-momentum by $p_\pi$. The subscripts distinguish only the kinematic assignments. The amplitude is symmetric under $K_1\leftrightarrow K_2$. We define $s=M^2=(p_1+p_2+p_\pi)^2$ and the pair invariants $t=(p_1+p_\pi)^2$, $u=(p_2+p_\pi)^2$, and $v=(p_1+p_2)^2$, corresponding to the $K_1\pi$, $K_2\pi$, and $K_1K_2$ pairs. They satisfy $t+u+v=s+2M_K^2+M_\pi^2$. For a pair with invariant mass squared \(x\), daughter masses \(M_1,M_2\), and spectator mass \(M_3\), we define
\begin{align}
q(s,x;M_3)&=\frac{1}{2\sqrt s}\sqrt{\lambda(s,x,M_3^2)},\nonumber\\
k(x;M_1,M_2)&=\frac{1}{2\sqrt x}\sqrt{\lambda(x,M_1^2,M_2^2)},
\end{align}
with $\lambda(a,b,c)=(a-b-c)^2-4bc$ the K\"all\'en function. Here, $q$ is the momentum of the pair (or spectator) in the parent rest frame, and $k$ is the momentum of either daughter in the pair rest frame.

Since the pair daughters and the spectator are pseudoscalars, a pair partial wave of orbital angular momentum $L$ has spin $L$. Coupling it to a parent with $J^P=0^-$ requires the orbital angular momentum between the pair and spectator to be $L$ as well. We retain $L=0,1,2$ and use the dimensionless angular vertices
\begin{align}
\mathcal V_L(q,k,z)&=\left(\frac{qk}{\Lambda^2}\right)^L
P_L(z)B_L(q)B_L(k)
\label{eq:vertices}
\end{align}
where $P_L$ is a Legendre polynomial, $z$ is the helicity-angle cosine defined below, and $\Lambda=1$~GeV sets the normalization convention. The factors $q^L$ and $k^L$ give the threshold behavior at the parent and pair vertices, respectively. We use the Blatt--Weisskopf barrier factors~\cite{Blatt:1952ije,VonHippel:1972fg}, normalized to $B_L(0)=1$,
\begin{align}
B_0(p)&=1,\nonumber\\
 B_1(p)&=\frac{1}{\sqrt{1+p^2R^2}},\nonumber\\
B_2(p)&=\frac{3}{\sqrt{9+3p^2R^2+p^4R^4}}.
\end{align}
A common radius $R=3$~GeV$^{-1}$ is fixed as a phenomenological input for both vertices.

For the $K_1\pi$ pair with spectator $K_2$,  we define $q_t=q(s,t;M_K)$ and $k_t=k(t;M_K,M_\pi)$. The cosine of the angle between the pion and $K_2$ in the $K_1\pi$ rest frame is
\begin{align}
z_t&=\frac{u_0(s,t)-u}{2q_tk_t}\sqrt{\frac{t}{s}},
\end{align}
with $u_0(s,t)=M_K^2+M_\pi^2+
{\left(t+M_\pi^2-M_K^2\right)\left(s-t-M_K^2\right)}/{2t}$.
The corresponding vertices are
\begin{align}
V(s,t,u)&=\mathcal V_1(q_t,k_t,z_t),\nonumber\\
D(s,t,u)&=\mathcal V_2(q_t,k_t,z_t).
\end{align}
For the $K_2\pi$ pair with spectator $K_1$, the same angle convention gives $V(s,u,t)$ and $D(s,u,t)$ by interchanging $t$ and $u$. For the $K_1K_2$ pair with spectator pion, define $q_v=q(s,v;M_\pi)$ and $k_v=k(v;M_K,M_K)$. The cosine of the angle between $K_2$ and the pion in the $KK$ rest frame and the associated tensor vertex are
\begin{align}
z_v=\frac{t-u}{4q_vk_v}\sqrt{\frac{v}{s}},\quad 
D_{KK}=\mathcal V_2(q_v,k_v,z_v).
\end{align} 

The fixed Breit--Wigner pair shapes are
\begin{align}
b_R(x)&=\frac{\Lambda}{M_R^2-x-i M_R^{}\Gamma_R^{} r_L(x)},
\label{eq:bw}
\end{align}
with the energy dependence of the width given by
\begin{align}
r_L(x)&=\left[\frac{k(x)}{k(M_R^2)}\right]^{2L+1}
\left[\frac{B_L(k(x))}{B_L(k(M_R^2))}\right]^2.
\end{align}
The fixed parameters are listed in Table~\ref{tab:pair}, with reference values from the PDG~\cite{ParticleDataGroup:2026mpi}. For $K_0^*(1430)$, the PDG central mass and width give a visibly poorer description of the two-dimensional $K\pi$ mass distribution. We therefore scanned its mass and width within their quoted uncertainties and selected $M_R=1457$~MeV and $\Gamma_R=210$~MeV, which improves the description. These values are held fixed in the subsequent fits.

\begin{table}[t]\centering
\caption{Mass and width parameters for the Breit--Wigner shapes, which are fixed in the fit. }\label{tab:pair}
\begin{tabular}{lrr}\toprule
Resonance &$M_R$ (MeV)&$\Gamma_R$ (MeV)\\\midrule
$K^*(892)$&896&47\\
$K^*(1410)$&1414&232\\
$K_0^*(1430)$&1457&210\\
$K_2^*(1430)$&1432&109\\
$K_0^*(1950)$&1957&170\\
$a_0(1710)$&1713&107\\
$a_2(1320)$&1318&107\\
\bottomrule
\end{tabular}
\end{table}

The scalar $KK$ sector contains two fixed source functions $a_1(v)=T^a_{21}(v)/\Lambda$ and $a_2(v)=T^a_{22}(v)/\Lambda$ from a two-channel $(\pi\eta,KK)$ implementation adapted from the chiral model of Ref.~\cite{Albaladejo:2015aca}, with the $\delta_{12}=100^\circ$ parameter set. They describe the same interacting scalar subsystem rather than two independent Flatt\'e resonances. We retain the analytic source functions up to the pair invariant mass $\sqrt{v}=2.50$~GeV. Between $2.50$ and $2.52$~GeV, each source is replaced by a cubic Hermite polynomial that matches its value and first derivative with respect to $\sqrt{v}$ at $2.50$~GeV and brings both to zero at $2.52$~GeV. The sources vanish above this endpoint. This prescription suppresses the high-energy extension of the scalar sources while keeping their values and first derivatives continuous. It is a fixed model cutoff, and none of the scalar scattering parameters are fitted here.

For brevity, we define $[b]_{t+u}=b(t)+b(u)$, $[bV]_{t+u}=b(t)V(s,t,u)+b(u)V(s,u,t)$, and similarly for $[bD]_{t+u}$. These combinations enforce kaon exchange symmetry. Table~\ref{tab:basis} lists the twelve basis functions $f_i(s,t,u)$, including smooth $K\pi$ $S$-, $P$-, and $D$-wave contributions. 

\begin{table}[t]\centering
\caption{Twelve independent basis functions. }\label{tab:basis}
\begin{tabular}{cll}\toprule
$i$&channel &$f_i(s,t,u)$\\\midrule
1,2&$a_0(980)\pi$&$a_1(v),\ a_2(v)$\\
3&\(K^*(892)\bar K\)&$[b_{K^*(892)}V]_{t+u}$\\
4&\(K_0^*(1430)\bar K\)&$[b_{K_0^*(1430)}]_{t+u}$\\
5&\(a_2(1320)\pi\)&$b_{a_2}(v)D_{KK}$\\
6&\(K_2^*(1430)\bar K\)&$[b_{K_2^*(1430)}D]_{t+u}$\\
7&\(a_0(1710)\pi\)&$b_{a_0(1710)}(v)$\\
8&\(K_0^*(1950)\bar K\)&$[b_{K_0^*(1950)}]_{t+u}$\\
9&\(K^*(1410)\bar K\)&$[b_{K^*(1410)}V]_{t+u}$\\
10&smooth $(K\pi)_S\bar K$ &$2/\Lambda$\\
11&smooth $(K\pi)_P\bar K$&$[V(s,t,u)+V(s,u,t)]/\Lambda$\\
12&smooth $(K\pi)_D\bar K$&$[D(s,t,u)+D(s,u,t)]/\Lambda$ \\\bottomrule
\end{tabular}
\end{table}

\section{Spectral density}\label{app:spectral}
Finite-width effects and interference enter the spectral density through phase-space integrals of the full pair shapes. For labeled kinematics,
\begin{equation}
\dd\Phi_3=\frac{q(M^2,x;M_3)k(x;M_1,M_2)}{64\pi^3M\sqrt{x}}\dd x\dd z\,.
\end{equation}
For the twelve shapes listed in Table~\ref{tab:basis}, the spectral density is
\begin{align}
\rho_{ij}(s)=6\times\frac12\int\dd\Phi_3\,
f_i(s,t,u)^*f_j(s,t,u),
\label{eq:rho}
\end{align}
for $i,j=1,\ldots,12$. The factor $1/2$ accounts for the identical kaons, and the shapes include the coherent sum over the two $K\pi$ pairings. All overlaps are evaluated, including those between different pair partial waves and pairings. No block-diagonal restriction is imposed. Thus both the dispersive loop and the predicted distributions retain interference between the different pair configurations. The overall factor 6 implements the adopted effective charge-multiplicity convention.

The coupling restrictions explain the parameter count in Table~\ref{tab:count}. We number the seven bare parent states by $\alpha=1,\ldots,7$ in the order introduced in Sec.~\ref{sec:formalism}, and assign $\alpha=8$ to the additional bare state in $H_1$. With the component ordering of Table~\ref{tab:basis}, the bare state $\alpha=1$ couples to components $\{1,2,6\}$, the states $\alpha=2,3$ to $\{1,2,3,5,6\}$ and the state $\alpha=4$ to all components except 8. States $\alpha=5,6,7$, and the additional state $\alpha=8$ in $H_1$, couple to all twelve components. These restrictions apply only to the direct couplings $g_{i\alpha}$. The model includes the full symmetric interaction $B$ and all spectral overlaps.

\begin{table}[t]\centering
\caption{Summary of real fit parameters. }\label{tab:count}
\begin{tabular}{lrr}\toprule
Parameter group&$H_0$&$H_1$\\\midrule
Bare squared masses $M_{0\alpha}$&7&8\\
Absorption strengths $\Gamma_{0\alpha}$&7&8\\
Allowed real couplings $g_{i\alpha}$&60&72\\
Symmetric interaction $B$&78&78\\
Complex sources $p^{(0)},p^{(1)}$&48&48\\
Bare production $\beta$, phase fixed&13&15\\
\midrule
Total&213&229\\\bottomrule
\end{tabular}
\end{table}

\section{Parent-mass efficiency approximation}\label{app:efficiency}
We construct an effective one-dimensional response from the reconstructed phase-space (PHSP) Monte Carlo histogram in Fig.~2(c) of Ref.~\cite{BESIII:2026rzt}. The simulated process is generated uniformly in four-body phase space, with the parent-mass distribution given by
\begin{align}
g_{\rm PHSP}(M)&\propto 2M\left(1-\frac{M^2}{M_{J/\psi}^2}\right)\Phi_3(M),\\
\Phi_3(M)&\propto\int_{(M_K+M_\pi)^2}^{(M-M_K)^2}\!\dd x\,
\frac{\sqrt{\lambda(M^2,x,M_K^2)}}{M^2}
\frac{\sqrt{\lambda(x,M_K^2,M_\pi^2)}}{x}.
\end{align}
For each 30~MeV bin, let $R_a$ be the extracted reconstructed PHSP height and $G_a$ the integral of $g_{\rm PHSP}$ over that bin. Normalizing both over the set $\mathcal V$ of 57 resolvable bins gives
\begin{equation}
r_a=\frac{R_a/\sum_{b\in\mathcal V}R_b}
{G_a/\sum_{b\in\mathcal V}G_b}.
\end{equation}

\begin{figure}[!tbp]
\centering\includegraphics[width=\linewidth]{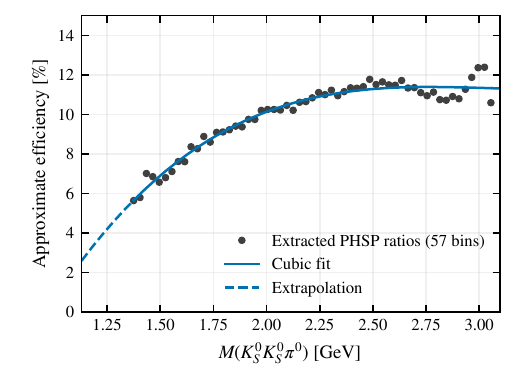}
\caption{Approximate parent-mass efficiency. Points are the resolved PHSP ratios after the reference-scale calibration. The solid curve is the cubic fit, and dashed segments show its extrapolation to the physical endpoints. No experimental uncertainty band is available.}\label{fig:efficiency}
\end{figure}

For a reference scale, we use $\varepsilon_a=C_\varepsilon r_a$ with $C_\varepsilon=10.8\%$. This approximate calibration combines the signal curve of Fig.~2(d) in Ref.~\cite{BESIII:2026rzt}, integrated over 1.95--2.75~GeV, with the product branching fraction $3.25\times10^{-4}$ from Table~I of Ref.~\cite{BESIII:2026mvn}.
A constant rescaling $\varepsilon\to c\varepsilon$ can be absorbed exactly into the fit parameters of our model and supplies no independent constraint on the scattering amplitude. The extracted efficiencies are shown in Fig.~\ref{fig:efficiency}, together with the cubic polynomial fit (blue curve), given by
\begin{align}
\varepsilon(M)=0.10+0.041x-0.042x^2+0.013x^3,
\label{eq:efficiency}
\end{align}
with $x=M/\mathrm{GeV}-2$.

\section{Two-dimensional distributions}\label{app:numerical}
\begin{figure*}[!tp]
\centering\includegraphics[width=\textwidth]{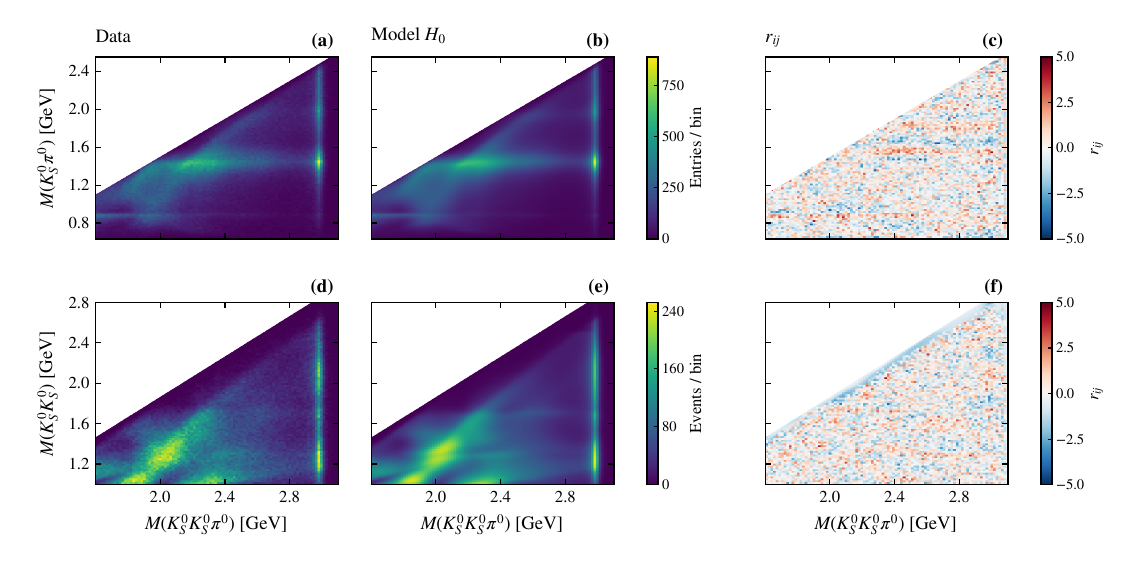}
\caption{Two-dimensional comparison for the efficiency-corrected $H_0$ fit. The upper row shows the $K_S^0\pi^0$ mass against the three-body mass, and the lower row shows the $K_S^0K_S^0$ mass against the three-body mass. The columns display digitized data, model predictions, and the signed deviance residuals $r_{ij}$ of Eq.~\eqref{eq:rij}, respectively.}\label{fig:h0}
\end{figure*}

\begin{figure*}[!tp]
\centering\includegraphics[width=\textwidth]{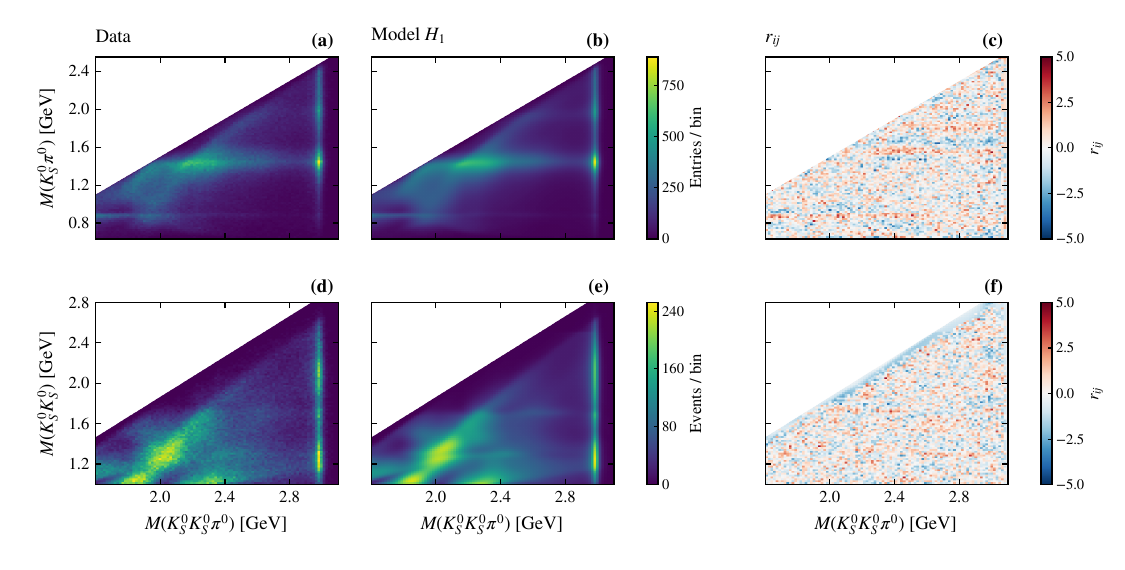}
\caption{The same two-dimensional comparison as in Fig.~\ref{fig:h0} for the efficiency-corrected $H_1$ fit.}\label{fig:h1}
\end{figure*}

Figures~\ref{fig:h0} and~\ref{fig:h1} compare the digitized two-dimensional distributions with the $H_0$ and $H_1$ predictions. Both predictions use the same parent-mass efficiency and fitted parameters as the projections in Fig.~\ref{fig:projections}. The maps show the signed square root of the per-bin Poisson deviance,
\begin{equation}
r_{ij}^{(r)}=\operatorname{sgn}\left(D_{ij}^{(r)}-\mu_{ij}^{(r)}\right)\sqrt{d_{ij}^{(r)}}.\label{eq:rij}
\end{equation}
Kinematically forbidden regions are blank. Positive values indicate that the model underestimates the reconstructed intensity.

\section{Coherent decomposition of the parent-mass spectrum}\label{app:decomposition}
\begin{figure}[thbp]
\centering
\includegraphics[width=\linewidth]{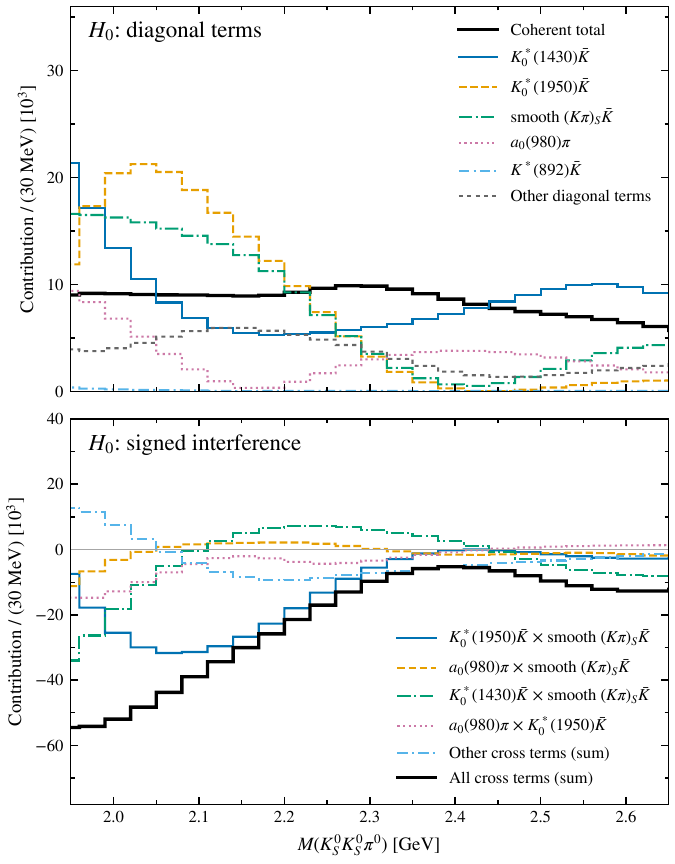}
\caption{Coherent decomposition of the parent-mass spectrum for the nominal $H_0$ fit. The upper panel shows the main diagonal contributions and coherent total. The lower panel shows the dominant signed interference terms and their sum. All terms include the same parent-mass efficiency.}\label{fig:h0-decomposition}
\end{figure}

{
To identify the contributions that build up the enhancement near $2.37\,\mathrm{GeV}$, Fig.~\ref{fig:h0-decomposition} decomposes the nominal $H_0$ spectrum into diagonal terms and signed interference terms. All terms are evaluated with the fitted amplitude of Eq.~\eqref{eq:threebody}, the same parent-mass efficiency, and the same normalization as the total spectrum. The two scalar source functions contributing to $a_0(980)\pi$ are combined coherently before this decomposition.

The $K_0^*(1430)\bar K$ diagonal contribution rises across the enhancement region, whereas the $K^*(892)\bar K$ contribution is small. Integrated over $2.0<M<2.4\,\mathrm{GeV}$, these two contributions amount to approximately $74\%$ and $1\%$ of the coherent total, respectively. The sum of all diagonal contributions is $371\%$, compensated by a net interference contribution of $-271\%$. The largest negative integrated cross term, between $K_0^*(1950)\bar K$ and the smooth $(K\pi)_S\bar K$ component, contributes $-195\%$. Therefore, the enhancement in this $H_0$ solution arises from the energy dependence and interference of the quasi-two-body amplitudes, without an additional bare state near $2.37\,\mathrm{GeV}$. Its line shape cannot be assigned to a single component. The resonant components are motivated by the two-body projections, their line shapes are fixed in the parent-amplitude fit, and the parent $K$-matrix background contains only a constant term. Nevertheless, the available mass projections need not constrain every parameter combination. This decomposition explains one viable solution, rather than establishing its uniqueness or excluding a resonance. A more precise determination requires a simultaneous analysis of multiple final states with full phase-space information, including angular distributions.
}

\bibliographystyle{elsarticle-num}
\bibliography{references}

\end{document}